\documentclass[fleqn,usenatbib]{iau_FM}

\usepackage{graphicx}
\usepackage{sidecap}
\usepackage{graphicx}
\usepackage{natbib}
\usepackage{multicol}
\usepackage{color}
\usepackage[usenames,dvipsnames]{xcolor}
\usepackage{url}
\usepackage{hyperref}
\usepackage{color}
\definecolor{darkblue}{rgb}{0.0, 0.0, 0.5}
\definecolor{darkred}{rgb}{0.8, 0.0, 0.0}

\hypersetup{colorlinks=true,citecolor=darkblue,linkcolor=darkblue}

\usepackage{setspace}

\def\aj{AJ}%
\def\apj{ApJ}%
\def\aap{A\&A}%
\def\mnras{MNRAS}%
\def\pasp{PASP}%
\renewenvironment{thebibliography}[1]{%
 \thebib@list
}{
 \endlist
}

\def\thebib@list{%
 \list{\null}{%
 \partopsep 0mm
 \leftmargin 1.2em
 \labelsep 0mm
 \itemindent -1.2em
 \itemsep 0.1\baselineskip
 \parsep 0mm
  \usecounter{enumi}%
 }%
}%

\title[STARSMOG] 
{What can the Occult do for you? \\ 
STarlight Attenuation \& Reddening Survey of Multiple Occulting Galaxies (STARSMOG)}

\author[Benne W. Holwerda \&William C. Keel]   
{B.W. Holwerda$^1$
 \and W.C. Keel$^2$
 }

\affiliation{$^1$Leiden Observatory \\ email: {\tt holwerda@strw.leidenuniv.nl} twitter: {\tt @benneholwerda}\\[\affilskip]
$^2$Alabama \\email: {\tt keel@ua.edu} twitter: {\tt @NGC3314}}

\pubyear{2015}
\jname{Astronomy in Focus, Volume 1} 
\editors{C. Leitherer, ed.}
\begin{document}

\maketitle

\begin{abstract}
Interstellar dust is still the dominant uncertainty in Astronomy, limiting precision in e.g., cosmological distance estimates and models of how light is re-processed within a galaxy. When a foreground galaxy serendipitously overlaps a more distant one, the latter backlights the dusty structures in the nearer foreground galaxy. Such an overlapping or occulting galaxy pair can be used to measure the distribution of dust in the closest galaxy with great accuracy. The STARSMOG program uses HST observation of occulting galaxy pairs to accurately map the distribution of dust in foreground galaxies in fine ($<$100 pc)
detail.
Furthermore, Integral Field Unit observations of such pairs will map the effective extinction curve in these occulting galaxies, disentangling the role of fine-scale geometry and grain composition on the path of light through a galaxy.

The overlapping galaxy technique promises to deliver a clear understanding of the dust in galaxies: the dust geometry, a probability function of the amount of dimming as a function of galaxy type, its dependence on wavelength, and evolution of all these properties with cosmic time using distant, high-redshift pairs.
\end{abstract}

\begin{figure*}[h]
  \begin{center}	
        \begin{minipage}[rt]{0.55\linewidth}
	\includegraphics[width=\textwidth]{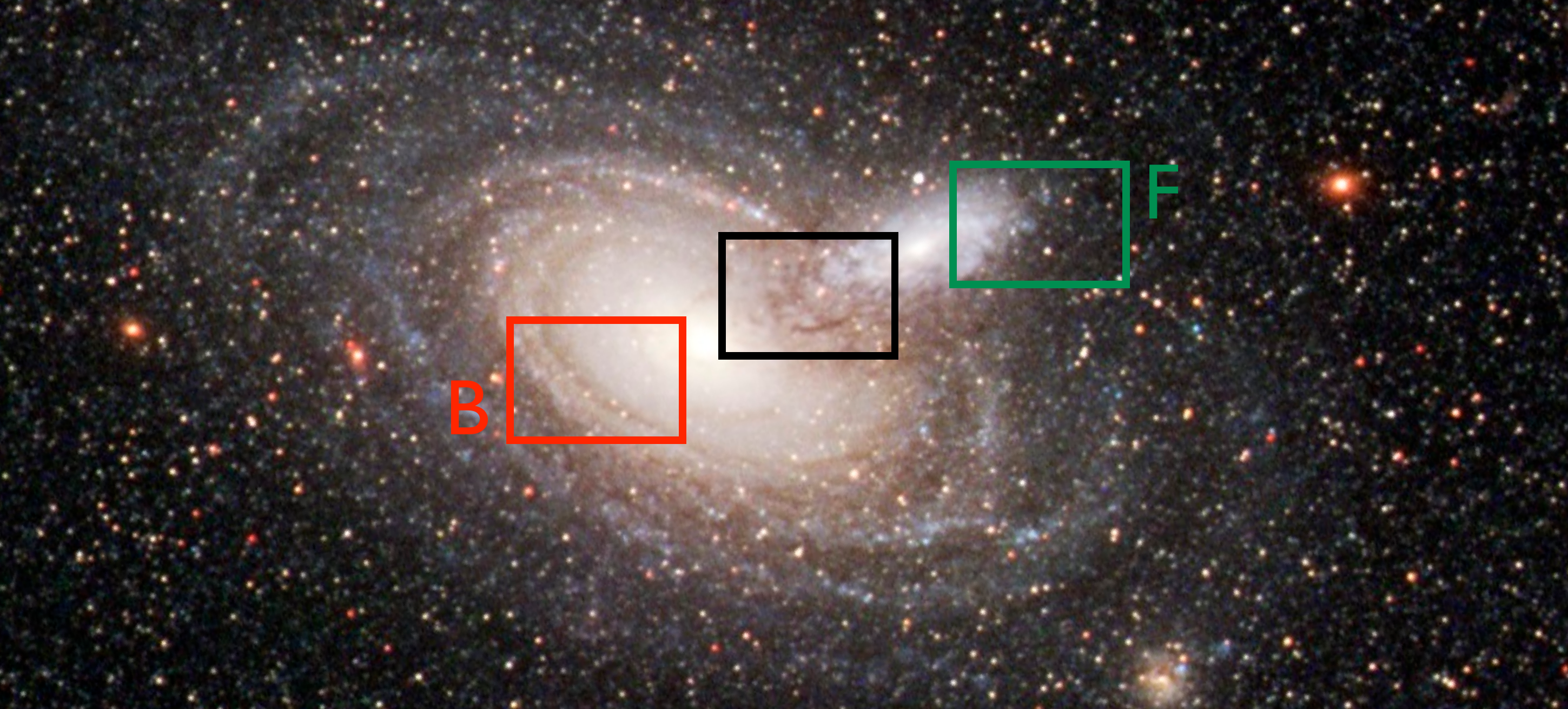}
	\caption{\label{f:pair}An occulting galaxy pair at z=0.06. The extinction in the overlap region ({\bf black aperture}) can be estimated from the complementary apertures; the foreground spiral ({\color{ForestGreen}F, the green aperture}) and the background galaxy ({\color{red}B, red aperture}).}
	\end{minipage}\hfill
	\begin{minipage}[rt]{0.44\linewidth}
    	\includegraphics[width=\textwidth]{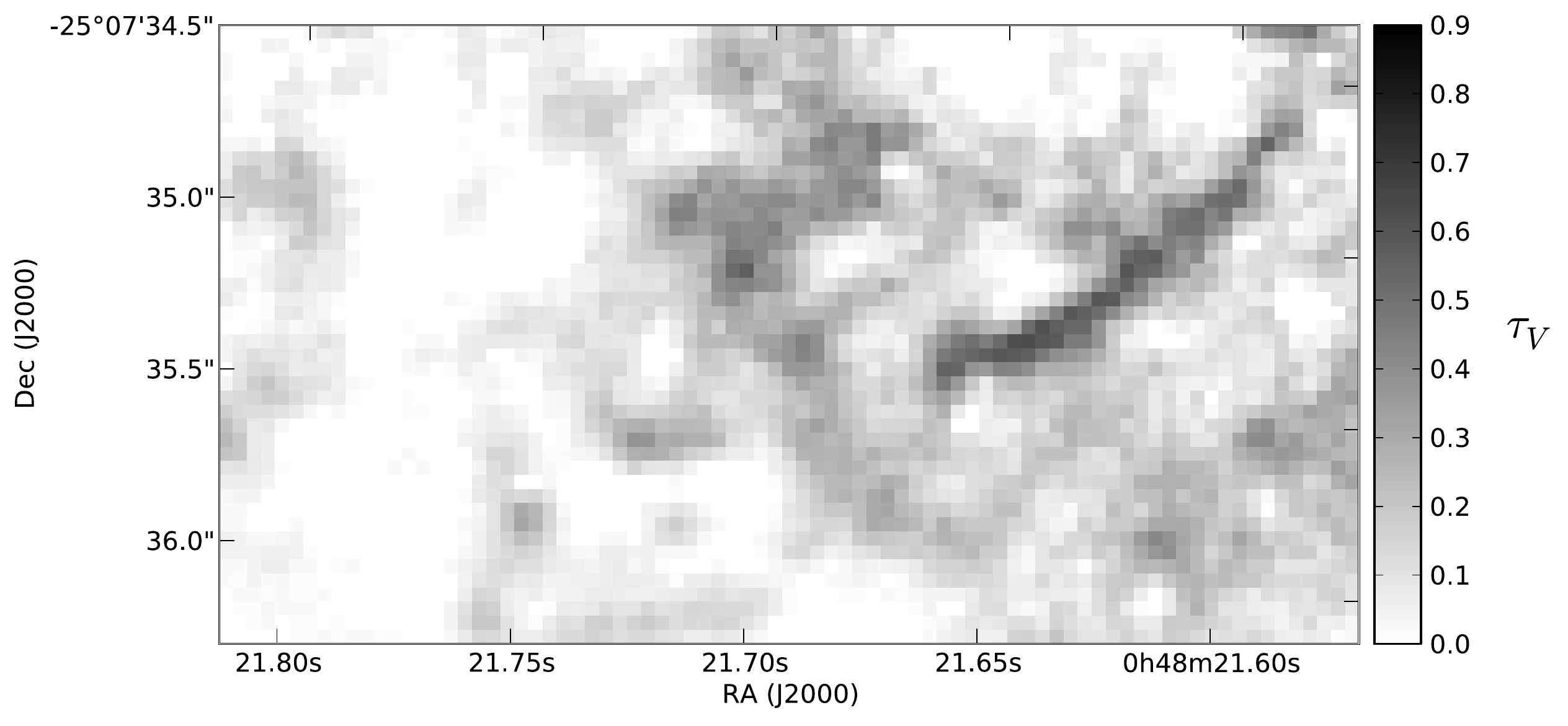}
	\caption{\label{f:taumap} The optical depth ($A_V$) map of the overlap region in Fig. \ref{f:pair}. The distribution of $A_V$ serves as the probability function of extinction --$P(A_V)$-- for a foreground galaxy of the mass and inclination at this radius.}
	\end{minipage}
  \end{center}
\end{figure*}

Interstellar dust is still the dominant astrophysical unknown in Cosmological distance estimates and models of how starlight is re-processed within a galaxy. 
When a galaxy accidentally overlaps a more distant one, the latter magnificently backlights the dusty structures in the nearest galaxy (Fig. \ref{f:pair}). Such an overlapping or {\em occulting} galaxy pair can be used to measure the distribution of dust in galaxies with great accuracy (Fig. \ref{f:taumap}).\\ 
The STARSMOG 
project uses HST observations of occulting galaxy pairs to map the fine-scale structure of dust extinction in galaxies to serve as a template for Astronomical observations \citep{Holwerda09,Keel14} and IFU observations to map the attenuation curves \citep[][]{Holwerda13a,Holwerda13b}.
%
%
%

Our motivation is twofold: first, the model for Universe today includes Dark Energy, inferred from the distances to Supernova. To evolve this technique to the next level of accuracy \citep[1\% in individual distances, see][]{Riess11}, an accurate model for the extinction properties of Supernova host galaxies will need to be developed \citep{DETF,Holwerda08a,Holwerda15a,Holwerda15c}. 
Secondly, 30\% of the light from stars in spiral galaxies is absorbed by the interstellar dust grains and re-emitted at longer 
wavelengths. Spectral Energy Distribution (SED) models need the distribution and detailed geometry of dust in galaxies \citep[e.g.,][]{Holwerda12a, Verstappen13, Allaert15}. 
%

The distribution of extinction values observed in an occulting pair can serve as an attenuation and reddening probability template, $P(A_V)$ and $P(R_V)$ respectively, for a SN\,Ia in a similar host galaxy as the foreground galaxy and both probability functions serve as a tight statistical constraint on any SED model of a spiral disk. 
\begin{figure*}[h]
  \begin{center}
          \begin{minipage}[rt]{0.6\linewidth}
    	\includegraphics[width=\textwidth]{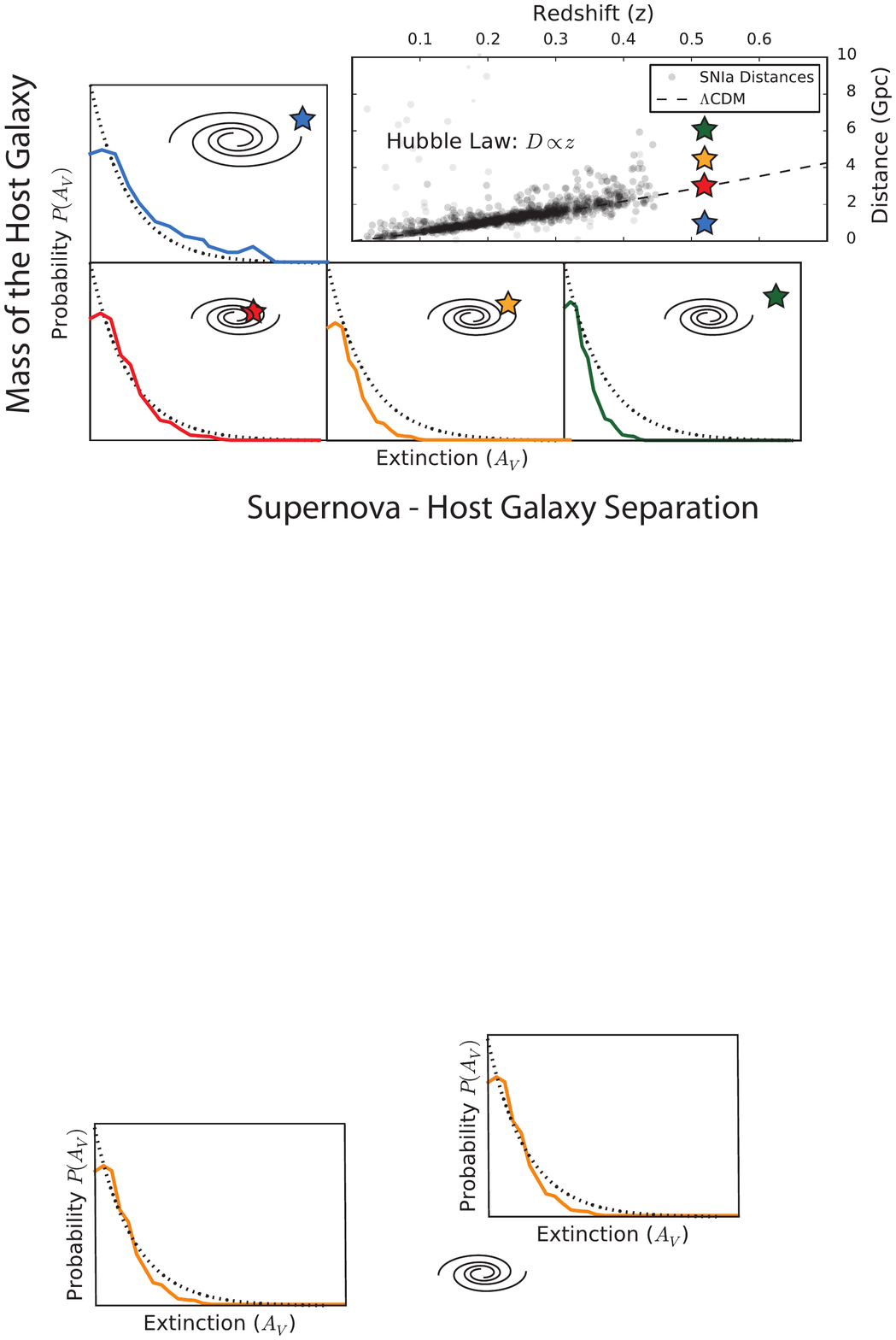}
	\end{minipage}\hfill
	\begin{minipage}[rt]{0.4\linewidth}
	\caption{\label{f:sketch} A sketch of the Hubble Law observed with supernovae (top right) with four cases of SNIa with different host galaxy extinction probabilities $P(A_V)$: (1) a massive galaxy: higher of dust extinction likely; an underestimate of actual distance, (2) a nominal galaxy; correct distance, (3) greater and greater supernova host separation; lower probability of dust extinction and over-estimation of distances.}
	\end{minipage}
  \end{center}
\end{figure*}


There are three ways to find overlapping pairs: visual identification \citep{Keel13}, blended spectra \citep{Holwerda07c,Holwerda15b}, or false close galaxy pairs in a highly complete spectroscopic survey \citep{Robotham14}. STARSMOG targets were selected using all three methods plus a $z<0.05$ requirement in order to map fine detail: 150 SNAP observations proposed, 49 (to date) executed with WFC3/F606W.


Overlapping galaxies offer the opportunity to map the dust attenuation and reddening in galaxy disks to great accuracy, providing probabilities for future high-precision studies of stellar populations and standard candle distance measurements.


\begin{multicols}{3}[]
\raggedcolumns
\addtocounter{unbalance}{1}
{\scriptsize

}
\end{multicols}

\end{document}